\documentclass[pra,aps,twocolumn,superscriptaddress,showpacs]{revtex4}

\usepackage{graphicx}

\usepackage{bm}

\begin{document}

\title{Gaussian localizable entanglement}

\author{Jarom\'{\i}r Fiur\'{a}\v{s}ek}
\affiliation{Department of Optics, Palack\'{y} University, 
17. listopadu 50, 77200 Olomouc, Czech Republic}

\author{Ladislav Mi\v{s}ta, Jr.}
\affiliation{Department of Optics, Palack\'{y} University, 
17. listopadu 50, 77200 Olomouc, Czech Republic}

\begin{abstract}
We investigate localization of entanglement of multimode Gaussian states into a pair of
modes by local Gaussian measurements on the remaining modes and classical communication. 
We find that for pure states and for mixed symmetric states maximum entanglement between
two modes can be localized by local homodyne detections, i.e. projections onto
infinitely squeezed states. We also show that non-Gaussian 
measurements allow to  localize more entanglement than Gaussian ones.

\end{abstract}


\maketitle

Quantum entanglement, the non-classical correlations exhibited by quantum systems,
lies at the heart of quantum information theory. 
Of particular interest are multipartite entangled states which provide a resource 
for one-way quantum computing \cite{Raussendorf01} and 
could form a backbone of the quantum communication network, where a part of the 
entangled system is located at each node of the network. For quantum communication 
purposes, it is of great interest to determine how much
entanglement can be localized on average between two nodes of the network 
by performing local measurements on the states located at the remaining nodes of the
network and announcing the measurement results to the two nodes. This so-called
localizable entanglement has been introduced and studied in the context of quantum 
spin chains \cite{Verstraete04,Popp05}.

In the present paper we investigate localization of entanglement in quantum
continuous-variable systems \cite{Braunstein05}. More specifically, we consider multimode 
Gaussian states and investigate how much entanglement can be localized between two modes by local
Gaussian measurements on the other modes. We prove that for pure Gaussian
states as well as for mixed symmetric Gaussian states the optimal strategy is 
to carry out a balanced homodyne detection (BHD) on each mode, i.e. to project it 
on infinitely squeezed state. Interestingly, we find that if we allow 
for non-Gaussian measurement strategies then we can localize more entanglement. 
The entanglement localization can be demonstrated experimentally with present-day
technology. Various multimode entangled Gaussian states of light can be generated by
combining single-mode squeezed states on an array of beam splitters
\cite{Yonezawa04} and highly efficient homodyne detectors are also available.

The entanglement properties of multimode Gaussian states have been investigated 
previously  \cite{Audenaert02,Wolf04} with particular focus on the symmetric Gaussian 
states invariant under arbitrary permutation of modes
\cite{Adesso04,Adesso05,Adesso06,Serafini05}. These latter states 
could be used to establish a continuous-variable quantum teleportation 
network \cite{vanLoock00,Yonezawa04} where quantum teleportation occurs between two (arbitrarily
chosen) modes $A$ and  $B$ 
and the other parties assist the teleportation by performing local measurements
$\mathcal{M}_j$
on their modes $C_j$ and sending the outcomes to the receiver $B$. 
Adesso and Illuminati \cite{Adesso05} determined the optimal multimode symmetric state that for 
a given total amount 
of squeezing maximizes fidelity of teleportation of coherent states from A to B 
and also logarithmic negativity 
of the effective two-mode state of $A$ and $B$ when each $\mathcal{M}_j$ is balanced
homodyne detection of $p$ quadrature. Here we prove for all mixed symmetric Gaussian
states that such local balanced homodyning is
optimal among all (not only local) Gaussian measurements 
and maximizes the
entanglement established between 
modes A and B by Gaussian measurements on modes $C_j$.

Consider $N$-mode  Gaussian state $\rho_{AB\bm{C}}$ shared among $N$ parties 
$A$, $B$, $C_j$, $j=1,\ldots, N-2$, with each party possessing a single mode. 
Parties $C_j$ attempt to increase the entanglement between $A$ and $B$ by making
\emph{local} Gaussian measurements and communicating the measurement outcomes 
to $A$ and $B$. By a Gaussian measurement on  mode $C_j$ we mean any measurement 
consisting of using auxiliary modes prepared in vacuum states, passive and active linear 
optics (beam splitters, phase shifters and squeezers) and BHD. Any such measurement 
can be described by the positive operator valued measure (POVM) 
\begin{equation}
\Pi_j(\alpha_j)= \frac{1}{\pi} D_j(\alpha_j)\Pi_j^0 D_j^\dagger(\alpha_j).
\label{Pialpha}
\end{equation}
Here $D_j(\alpha_j)=\exp(\alpha_j c_j^\dagger-\alpha_j^\ast c_j)$ denotes the displacement
operator, $\Pi_j^0$ is a density matrix of a (generally mixed) single-mode Gaussian 
state with covariance matrix $\gamma_{C_j}^M$ and zero displacement and $\alpha_j$ is a certain 
linear combination of the quadrature values measured by the BHDs. In particular, homodyne detection
on $C_j$ is recovered in the limit of infinitely squeezed state $\Pi_j^0$. A crucial feature of 
the Gaussian measurement is that $\gamma_{C_j}^M$ depends only on the structure of the
linear optical network but not on the measurement outcomes of the BHDs.
The normalization $\mathrm{Tr}[\Pi_j^0]=1$  implies that
\begin{equation}
\frac{1}{\pi} \int D_j(\alpha_j) \Pi_j^0 D_j^\dagger(\alpha_j) d^2\alpha_j= \openone_j,
\label{completness}
\end{equation}
which ensures the completness of the POVM (\ref{Pialpha}).
Without loss of any generality we can assume that each $\Pi_j^0$ is a
projector onto a pure Gaussian state because any mixed Gaussian state can 
be expressed as a mixture of pure Gaussian states and classical mixing 
cannot increase entanglement.

The elements of the  total POVM describing measurement on all modes $C_j$
can be written as a product of the single-site elements,
$\Pi_{\bm C}(\bm{\alpha})=\bigotimes_{j=1}^{N-2} \Pi_{j}(\alpha_j)$, 
where $\bm{\alpha}=(\alpha_1,\cdots,\alpha_{N-2})$. 
The measurement outcome $\bm{\alpha}$ is obtained with
probability density $P(\bm{\alpha})=\mathrm{Tr}[\openone_{AB}\otimes \Pi_{\bm C}(\bm{\alpha})
\rho_{AB\bm{C}}]$ and the resulting normalized density matrix of the
conditional bipartite state shared by $A$ and $B$ reads 
\begin{equation}
\sigma_{AB}(\bm{\alpha})=\frac{1}{P(\bm{\alpha})}\mathrm{Tr}_{\bm{C}}[\openone_{AB}\otimes\Pi_{\bm C}(\bm{\alpha}) \,
\rho_{AB\bm{C}}].
\label{sigmaABz}
\end{equation}
Let $E[\sigma_{AB}]$ denotes a measure of entanglement 
of a bipartite state $\sigma_{AB}$.
We define the Gaussian localizable entanglement $E_{L,G}$ between $A$ and $B$ as 
the maximum entanglement that can be, on average, established between two parties $A$ and $B$  by 
local Gaussian measurements performed by $C_j$ on their modes and by communicating
the measurement outcomes to $A$ and $B$. We have
\begin{equation}
E_{L,G}=\max_{\Pi_{\bm C}}\int_{\bm{\alpha}} P(\bm{\alpha}) E[\sigma_{AB}(\bm{\alpha})]
\mathrm{d} \bm{\alpha},
\label{ELG}
\end{equation}
where the maximum is taken over all local Gaussian POVMs $\Pi_{\bm C}$.  

It follows from the properties of Gaussian operations
and measurements \cite{Giedke02} that for all measurement outcomes $\bm{\alpha}$ the conditionally
prepared state $\sigma_{AB}(\bm{\alpha})$ is a Gaussian state with fixed
covariance matrix and varying displacement which depends linearly on $\bm{\alpha}$. 
The entanglement properties of a Gaussian state
depend only on the covariance matrix, because the displacement can be set to zero by
means of suitable local displacement operations $D_j$. 
It holds that $E[\sigma_{AB}(\bm{\alpha})]=E[\sigma_{AB}(\bm{0})]$, $\forall \bm{\alpha}$
and from the definition (\ref{ELG}) it immediately follows that
 $E_{L,G}=\max_{\Pi_{\bm C}} E[\sigma_{AB}(\bm{0})]$. 
In order to determine the Gaussian localizable
entanglement it thus suffices to optimize over all local projections onto pure
single-mode squeezed vacuum states. Generally, this optimization is still a daunting
task and can be performed only numerically. However, we shall see that in
case of three-mode pure states a fully analytical expression can be derived.
Moreover, for general multimode pure Gaussian states we show that the optimal measurement involves
homodyne detection on each mode so it suffices to optimize over the phases $\theta_j$
which specify the quadratures being measured.

Consider a pure three-mode  Gaussian state $|\psi\rangle_{ABC}$ 
shared by parties $A$, $B$ and $C$ and characterized by a covariance 
matrix $\gamma_{ABC}$. After projection of mode $C$ onto a pure Gaussian
state, the modes $A$ and $B$ will be in a pure Gaussian state $|\phi\rangle_{AB}$ 
with covariance matrix
\begin{equation}
\gamma_{AB}=\left( \begin{array}{cc}
\gamma_A  & \delta_{AB} \\
\delta_{AB}^T & \gamma_B
\end{array}
\right).
\label{gammaAB}
\end{equation}
where $\gamma_{A}$ $(\gamma_B)$ denotes covariance matrix of mode  $A$ $(B)$  
and $\delta_{AB}$ contains correlations between the quadratures of the two modes.
A unique measure of entanglement of pure states is provided 
by the entropy of entanglement, which is the von Neumann entropy 
of the reduced density matrix of one party,  $E[|\phi\rangle_{AB}]
=-\mathrm{Tr}[\rho_A \log_2 \rho_A]$, where $\rho_A=\mathrm{Tr}_B[|\phi\rangle
\langle \phi|_{AB}]$. The von Neumann entropy of a single-mode Gaussian
state with covariance matrix $\gamma_A$ is a function of the symplectic invariant 
$n_{A}=(\sqrt{\det\gamma_{A}}-1)/2$. 
Explicitly, the entropy of entanglement reads
\begin{equation}
E[|\phi\rangle_{AB}]=(n_A+1)\log_2(n_A+1)-n_A\log_2(n_A).
\label{Ethermal}
\end{equation}
In order to maximize $E[|\phi\rangle_{AB}]$ we have to maximize $\det\gamma_A$.

\begin{figure}
\begin{center}
\includegraphics[width=\linewidth]{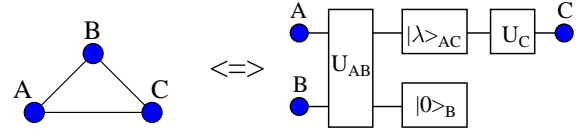}
\end{center}
\caption{Decomposition of pure three-mode Gaussian state.}
\label{equivalence_fig}
\end{figure}

We  prove that it is optimal 
to measure a suitably chosen quadrature of mode C $x_{C,\theta}=(c e^{-i\theta}+c^\dagger
e^{i\theta})/\sqrt{2}$.  Consider a bipartite splitting $AB|C$ of the pure state 
$|\psi\rangle_{ABC}$. It has been shown \cite{Holevo01} that  there exist
unitary Gaussian transformations  $\mathcal{U}_{AB}^\dagger$ and
$\mathcal{U}_{C}^\dagger$ 
acting on modes $AB$ and $C$,
respectively, which transform the three-mode pure Gaussian state $|\psi\rangle_{ABC}$ into a
product of a  two-mode squeezed vacuum state in modes $A$ and $C$ and a vacuum state 
in mode $B$,
$\mathcal{U}_{AB}^\dagger\otimes
\mathcal{U}_C^\dagger|\psi\rangle_{ABC}=|\lambda\rangle_{AC}|0\rangle_B$. 
Here
$|\lambda\rangle_{AC}=\sqrt{1-\lambda^2}\sum_{n=0}^\infty \lambda^n |n\rangle_A
|n\rangle_C$,
$\lambda^2=(\sqrt{\det\gamma_C}-1)/(\sqrt{\det\gamma_C}+1)$, $\gamma_C$ is CM of mode
$C$ prior to measurement and $|n\rangle$ denotes the $n$-photon Fock state.
The situation is illustrated in Fig.~\ref{equivalence_fig}. The transformation
$\mathcal{U}_C$ can be absorbed
into the Gaussian measurement on mode $C$. Projection of one part of the two-mode
squeezed vacuum state onto single-mode squeezed state with covariance matrix
$\tilde{\gamma}_C=W(\theta) V(r) W^T(\theta),$ where 
\begin{equation}
W(\theta)= \left(
\begin{array}{cc}
\cos\theta  & \sin \theta \\
-\sin\theta & \cos\theta 
\end{array}
\right),
\qquad
V(r)= \left(
\begin{array}{cc}
e^{2r}  & 0 \\
0 & e^{-2r} 
\end{array}
\right),
\label{UVdefinition}
\end{equation}
prepares the mode $A$ in a similar pure single-mode squeezed vacuum state 
$|s;\theta\rangle$ with covariance
matrix
$\tilde{\gamma}_A=W(-\theta) V(s) W^T(-\theta),$
where 
\begin{equation}
e^{2s}=\frac{1-\lambda^2+(1+\lambda^2)e^{2r}}{1+\lambda^2+(1-\lambda^2)e^{2r}}.
\label{s}
\end{equation}
The resulting two-mode Gaussian state of modes $A$ and $B$ can
be obtained from the product Gaussian state of modes $A$ and $B$ by the action of
$\mathcal{U}_{AB}$,
$|\phi\rangle_{AB}=\mathcal{U}_{AB}|s;\theta\rangle_A|0\rangle_B$.
Let $S_{AB}$ denotes the symplectic matrix corresponding to the unitary
$\mathcal{U}_{AB}$ which in the Heisenberg picture governs the linear transformation
of quadrature operators. We  decompose $S_{AB}$ with respect to the $A|B$ splitting,
\begin{equation}
S_{AB}= \left(
\begin{array}{cc}
S_{AA}  & T_{AB} \\
T_{BA} & S_{BB} 
\end{array}
\right).
\label{SABdefinition}
\end{equation}
The covariance matrix $\gamma_{AB}$ of the state $|\phi\rangle_{AB}$ can be
expressed as 
$\gamma_{AB}= S_{AB} (\tilde{\gamma}_A \oplus \openone_B) S_{AB}^T,$
where the identity matrix $\openone_B$ represents  the covariance matrix of
vacuum state. After a straightforward calculation we arrive at  
$\gamma_A=S_{AA}\tilde{\gamma}_A S_{AA}^T+T_{AB}T_{AB}^T.$
To calculate $\det\gamma_A$ which we want to maximize we  use the formula
for  determinant of a sum of two  $2\times 2$ symmetric matrices $X$ and $Y$,
\begin{equation}
\det(X+Y)=\det{X}+\det{Y}+\mathrm{Tr}[XRYR^T],
\label{sumdeterminant}
\end{equation}
where
$R= \left(
\begin{array}{cc}
0  & 1 \\
-1 & 0 
\end{array}
\right).$
Taking into account that determinant of a
covariance matrix of a pure Gaussian state is equal to unity,
$\det{\tilde{\gamma}_A}=1$, we arrive at
\begin{equation}
\det\gamma_A=(\det S_{AA})^2+(\det T_{AB})^2+\mathrm{Tr}[\tilde{\gamma}_A M],
\end{equation}
where $M=S_{AA}^T R T_{AB}T_{AB}^T R^T S_{AA}$ is a symmetric positive semidefinite matrix.
$M$ can be diagonalized by orthogonal rotation, 
$W(\theta_0)M W^T(\theta_0)=\mathrm{diag}(M_{xx},M_{pp})$.
Since we optimize over all phases $\theta$ we can, without loss of any generality,
make the substitution $\theta \mapsto \theta-\theta_0$ and assume that $M$ 
is diagonal. The nontrivial part of $\det\gamma_A$ which should be maximized then reads
\begin{equation}
e^{2s}(M_{xx}\cos^2\theta+M_{pp}\sin^2\theta)
+e^{-2s}(M_{xx}\sin^2\theta+M_{pp}\cos^2\theta).
\label{sthetaexpression}
\end{equation}
The maximum of this function should  be found in the interval 
$s\in[-s_{\mathrm{max}},s_{\mathrm{max}}]$, where
$e^{2s_{\mathrm{max}}}=(1+\lambda^2)/(1-\lambda^2)$, c.f. Eq. (\ref{s}). The maximum
 squeezing  $\pm s_{\mathrm{max}}$
is obtained if mode $C$ is projected onto the eigenstate of quadrature operator using
homodyne detection. It is straightforward to check that for any $\theta$ expression 
(\ref{sthetaexpression}) is maximized if $s=s_{\mathrm{max}}$ or $s=-s_{\mathrm{max}}$.
This proves that the optimal Gaussian measurement on C which maximizes the
entanglement between $A$ and $B$ is homodyne detection. This remains valid even if we
take into account the  squeezing
operation $\mathcal{U}_C$ which is included in this measurement. The only difference
is that instead of quadrature $q_{C}$ we should measure quadrature
$x_{C,\theta_{\mathrm{opt}}}$
defined as
$\mathcal{U}_C q_{C} \mathcal{U}_C^\dagger=a_x x_{C}+a_p
p_{C}=\sqrt{a_x^2+a_p^2}\,x_{C,\theta_{\mathrm{opt}}}$. 
The optimal phase $\theta_{\mathrm{opt}}$  specifying the quadrature
$x_{C,\theta_{\mathrm{opt}}}$ can be determined analytically by solving quadratic
equation for $\tan\theta_{\mathrm{opt}}$ \cite{Mista07}. Alternatively, the three-mode pure state can be
transformed by local unitary Gaussian operations to a standard form, where all
correlations between amplitude quadratures $x_j$ and phase quadratures $p_k$ vanish \cite{Adesso06}. In this case it is optimal to measure
either $x_C$ or $p_C$.

\begin{figure}[!t!]
\begin{center}
\includegraphics[width=0.8\linewidth]{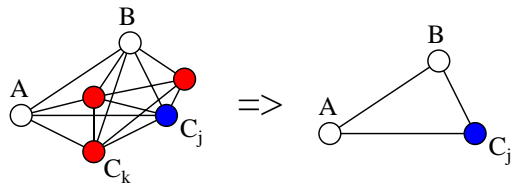}
\end{center}
\caption{Reduction of $N$-mode pure Gaussian state to three-mode pure Gaussian state
by local projections on modes $C_k$.}
\label{reduction_fig}
\end{figure}

We will next argue that for an arbitrary multimode pure Gaussian state
$|\psi\rangle_{AB\bm{C}}$ the maximum $E_{L,G}$ between $A$ and
$B$ is obtained if each party $C_j$ performs homodyne detection of some quadrature
$x_{C_j,\theta_j}$. The proof is based on the reduction argument. Consider the
multipartite state illustrated schematically in Fig.~\ref{reduction_fig}. Suppose that all parties
$C_k$ except for $C_j$ perform their local projections on pure Gaussian states
described by POVMs (\ref{Pialpha}).
This prepares the three-mode system $ABC_j$ in a pure Gaussian state with a fixed
covariance matrix that does not depend on the measurement outcomes $\alpha_k$ of
$C_k$, see Fig.~\ref{reduction_fig}.  For a three-mode pure Gaussian state we
proved that $E_{L,G}$ is maximized if $C_j$ makes a balanced
homodyne detection of appropriately chosen quadrature. We can thus see that,
irrespective of the measurements carried out by $C_k$, the optimal choice 
for $C_j$ is balanced homodyning. Note that $C_{j}$ even does not need to learn the
measurement outcomes $\alpha_k$ of the other parties $C_{k\ne j}$. 
It suffices if $A$ and $B$ receive from all parties $C_l$ the data
$\alpha_l$. By local displacement operations they can then deterministically 
compensate for the resulting displacements of their modes 
which are linear in $\alpha_l$. The above argument can be applied 
to any $C_k$ which proves that the optimal measurement  strategy must consist 
of balanced homodyning on each mode $C_k$.


We now consider general mixed symmetric Gaussian 
states  \cite{Adesso04,Adesso05} which are invariant under
arbitrary permutation of modes. This implies that the covariance matrix has 
a highly symmetric form,
\begin{equation}
\gamma_{\mathrm{sym}}=\left(
\begin{array}{cccc}
\beta & \epsilon & \ldots & \epsilon  \\
\epsilon & \beta &  \epsilon & \vdots  \\
\vdots & \epsilon & \ddots & \epsilon \\
\epsilon & \cdots & \epsilon & \beta
\end{array}
\right),
\label{gammasym}
\end{equation}
where $\beta$ and $\epsilon$ denote symmetric $2\times 2$ matrices.
By means of local canonical transformations it is possible to simultaneously diagonalize
both $\beta$ and $\epsilon$ \cite{Serafini05} so without loss of any generality we may assume that 
$\beta=\mbox{diag}(b,b)$ and $\epsilon=\mbox{diag}(\epsilon_{1},\epsilon_{2})$. 
We 
choose the logarithmic negativity $E_{\mathcal{N}}$ 
as the measure of entanglement. We prove that maximum  entanglement 
between two modes (labeled $A$ and $B$) can be localized  by homodyne detection of
either $x$ or $p$  on all remaining $N-2$ modes $C_j$. 
Consider the bipartite $2\times (N-2)$ splitting $AB|\bm{C}$. By means of unitary
Gaussian transformation on modes $C_j$ which can be physically realized by interference
on an array of $N-3$ unbalanced beam splitters all modes $C_k$, $k \neq 1$, 
can be decoupled from $A$,  $B$ and $C_1$ thereby effectively reducing the 
problem to three-mode case \cite{Serafini05}. Similarly, interference of $A$ with $B$ on a balanced beam
splitter decouples $B$ from $A$ and $C_1$. In this way we obtain the equivalent representation
shown in Fig.~\ref{equivalence_fig} where $\mathcal{U}_{AB}$ represents mixing on a balanced beam splitter, 
mode $B$ is initially in a mixed state with
$\gamma_{B,\mathrm{in}}=\beta-\epsilon$ while the initial CM of modes $A$ and $C_1$ reads
\begin{eqnarray}
\gamma_{AC_1,\mathrm{in}}=\left(
\begin{array}{cc}
\beta+\epsilon & \sqrt{2(N-2)} \epsilon \\
\sqrt{2(N-2)} \epsilon & \beta +(N-3)\epsilon
\end{array}
\right).
\label{gammaAC}
\end{eqnarray}
Projection of $C_1$ onto state with CM $\tilde{\gamma}_C$ prepares mode $A$ in 
state with CM
$\gamma_{A,{\rm in}}=\beta+\epsilon-2(N-2)\epsilon[\beta+(N-3)\epsilon+\tilde{\gamma}_C]^{-1}\epsilon$.
The logarithmic negativity $E_{\mathcal{N}}=\max(0,-\log_2 \mu)$ where $\mu$ is the
minimum symplectic eigenvalue of CM of partially transposed state of modes $A$ and $B$
\cite{Vidal02}.
After some algebra one finds that 
$\mu^2=\min[\mathrm{eig}(\gamma_{A,\mathrm{in}}R \gamma_{B,\mathrm{in}}R^T)]$
where $\mathrm{eig}(\mathcal{A})$ denotes eigenvalues of a matrix $\mathcal{A}$. 
In order to maximize $E_{\mathcal{N}}$ we have to minimize $\mu$ over all admissible $\tilde{\gamma}_C$.
This could be done analytically and one can prove \cite{Mista07} that it is optimal to measure either the
$x$ or $p$ quadrature of $C_1$, depending on the relation between $\epsilon_{1}$ and $\epsilon_2$. 
This optimal measurement is a joint measurement on the original modes $C_{j}$ that can be performed locally
by measuring either $x$ or $p$ quadrature of each mode and then properly averaging 
the results. $\mu^{2}$ is particularly simple for $N=3$ when it reads 
$\mu^{2}=(b-\epsilon_{1})(b-\epsilon_2)(1+2\mbox{min}(\epsilon_1,\epsilon_2)/b).$

\begin{figure}[!t!]
\begin{center}
\includegraphics[width=\linewidth]{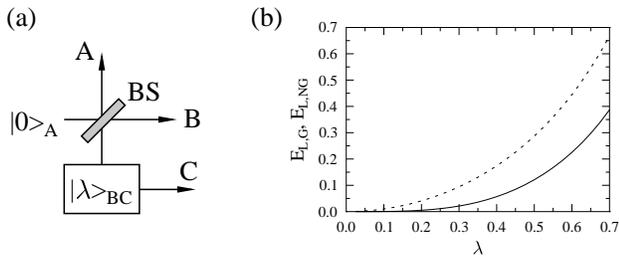}
\end{center}
\caption{(a) Scheme for preparation of a three-mode Gaussian state. (b) Localizable
entanglement $E_{L,G}$ (solid curve) and $E_{L,NG}$ (dashed curve) are 
plotted versus the squeezing parameter $\lambda$.}
\label{nonGaussian_fig}
\end{figure}


Finally, we show that if we allow also non-Gaussian measurements on modes ${\bm C}$ 
then we can localize a higher amount of entanglement between modes $A$ and $B$ 
than one can localize by Gaussian measurements. This is yet another example 
of the well known fact that non-Gaussian operations are optimal for
certain tasks such as cloning \cite{Cerf05} or partial estimation of coherent states \cite{Mista06}. 
We demonstrate the superiority of non-Gaussian measurements on a particular
illustrative example. Consider the state preparation scheme shown in Fig.~\ref{nonGaussian_fig}(a). 
The three-mode state is generated from a two-mode squeezed vacuum state
$|\lambda\rangle_{BC}$ in modes $B$ and $C$ by mixing mode $B$ on a balanced 
beam splitter BS with mode $A$ which is initially in a vacuum state. 

The optimal Gaussian measurement on $C$ which maximizes entanglement 
between $A$ and $B$ is a homodyne detection and
the Gaussian localizable entanglement $E_{L,G}$ can be evaluated 
from formula (\ref{Ethermal}) with
$n_A=\frac{1}{2}\sqrt{\frac{1}{1-\lambda^4}}-\frac{1}{2}.$
Suppose now that we would measure the number of photons $n$ in mode $C$. With
probability $p_n=(1-\lambda^2)\lambda^{2n}$ we would prepare in mode $B$ $n$-photon
Fock state $|n\rangle$, which then impinges on a balanced beam splitter BS, 
c.f. Fig.~\ref{nonGaussian_fig}(a). The resulting state of $A$ and $B$ expressed 
in the Fock state basis reads,
$|\psi_n\rangle_{AB}= \frac{1}{2^{n/2}}\sum_{k=0}^n \sqrt{n \choose k}
|k,n-k\rangle_{AB}.$
The entropy of entanglement of $|\psi_n\rangle_{AB}$ is given by
\begin{equation}
S_n=-\frac{1}{2^n}\sum_{k=0}^n {n\choose k} \log_2\left[\frac{1}{2^n} {n
\choose k}\right],
\end{equation}
and has to be evaluated numerically. The average entanglement between $A$ and $B$ 
is then $E_{L,NG}=\sum_{n=0}^{\infty}p_n S_n$. Both $E_{L,G}$ and $E_{L,NG}$ are plotted in
Fig.~\ref{nonGaussian_fig}(b) as a function of the squeezing parameter $\lambda$. We can see that
$E_{L,NG}> E_{L,G}$ so the non-Gaussian measurement strategy outperforms 
the best Gaussian one.

We acknowledge financial support from the Ministry of Education of the 
Czech Republic (LC06007 and MSM6198959213) and from the EU under project 
COVAQIAL (FP6-511004) and SECOQC (IST-2002-506813).

\end{document}